\documentclass[12pt,a4paper]{article}
\usepackage[utf8]{inputenc}
\usepackage[T1]{fontenc}
\usepackage{amsmath,amssymb}
\usepackage{graphicx}
\usepackage{caption,subcaption}
\label{Sec:Introduction}

\begin{document}
\textwidth=135mm
 \textheight=200mm
\begin{center}
{\bfseries Clusters in nuclear matter and Mott points
\footnote{{\small Talk at the 32th Max-Born-Symposium and HECOLS workshop on "Three Days of Phase Transitions in Compact Stars, Heavy-Ion Collisions and Supernovae", Institute for Theoretical Physics, University of Wroc\l{}aw, Wroc\l{}aw, Poland, February 17--19, 2014.}}}
\vskip 5mm
G. R\"{o}pke$^*$
\vskip 5mm
{\small {\it $^*$ Institut f\"{u}r Physik, Universit\"{a}t Rostock, D-18051 Rostock, Germany}}\\
\end{center}
\vskip 5mm
\centerline{\bf Abstract}
Light clusters (mass number $A \leq 4$) in nuclear matter at subsaturation densities are described using a quantum statistical 
approach. In addition to self-energy and Pauli-blocking, 
effects of continuum correlations are taken into account to 
calculate the quasiparticle properties and abundances of light elements. 
Medium-modified quasiparticle properties are important ingredients to derive a nuclear matter equation of state 
applicable in the entire region of warm dense matter below saturation density.
The influence of the nucleon-nucleon interaction on the quasiparticle shift is discussed.
\vskip 10mm

\section{\label{sec:intro}Introduction}

The nuclear matter equation of state (EOS) is an important ingredient to understand the properties of nuclear systems,
for instance for heavy ion collisions (HIC), see \cite{Natowitz}, 
and for astrophysical applications such as the formation of neutron stars in core-collapse supernovae, see \cite{Tobias}.
In both cases, matter is considered at subsaturation density (baryon density 
$n_B \leq n_{\rm sat} \approx 0.16$ fm$^{-3}$), moderate temperatures $T \leq 20$ MeV, 
and proton fraction $Y_p=n^{\rm tot}_p/n_B$ between 0
and 1 where $n^{\rm tot}_p$ denotes the total proton number density (including 'bound' and 'free' protons).

In this warm dense matter (WDM) region, correlations and bound state formation are relevant. We are considering the 
influence of the formation of light clusters (deuteron $d=^2\!\!{\rm H}$, triton $t=^3\!\!{\rm H}$, helion  $h=^3\!\!{\rm He}$, and  $\alpha=^4\!\!{\rm He}$)
on the EOS. Whereas at low densities the nuclear statistical equilibrium (NSE) is a reasonable approximation,
where non-interacting bound states in chemical equilibrium are considered so that the composition is given by a mass-action law
(continuum contributions are neglected), with increasing density the bound states are dissolved due to Pauli blocking 
and do not contribute to the EOS near to the saturation density.

About thirty years ago \cite{RMS,RMS2} a quantum statistical (QS) approach to the EOS has been given which 
allows for a systematic description of correlations in nuclear matter. There, the dissolution of bound states because of 
Pauli blocking has been denoted as Mott effect, and the phase diagram shown in Fig.~1 has been proposed.
Meanwhile, various theoretical investigations have been performed as well as first experimental evidences \cite{Natowitz}.

\begin{figure}[!hbt]
\begin{subfigure}{0.58\textwidth}
  \includegraphics[width=\textwidth]{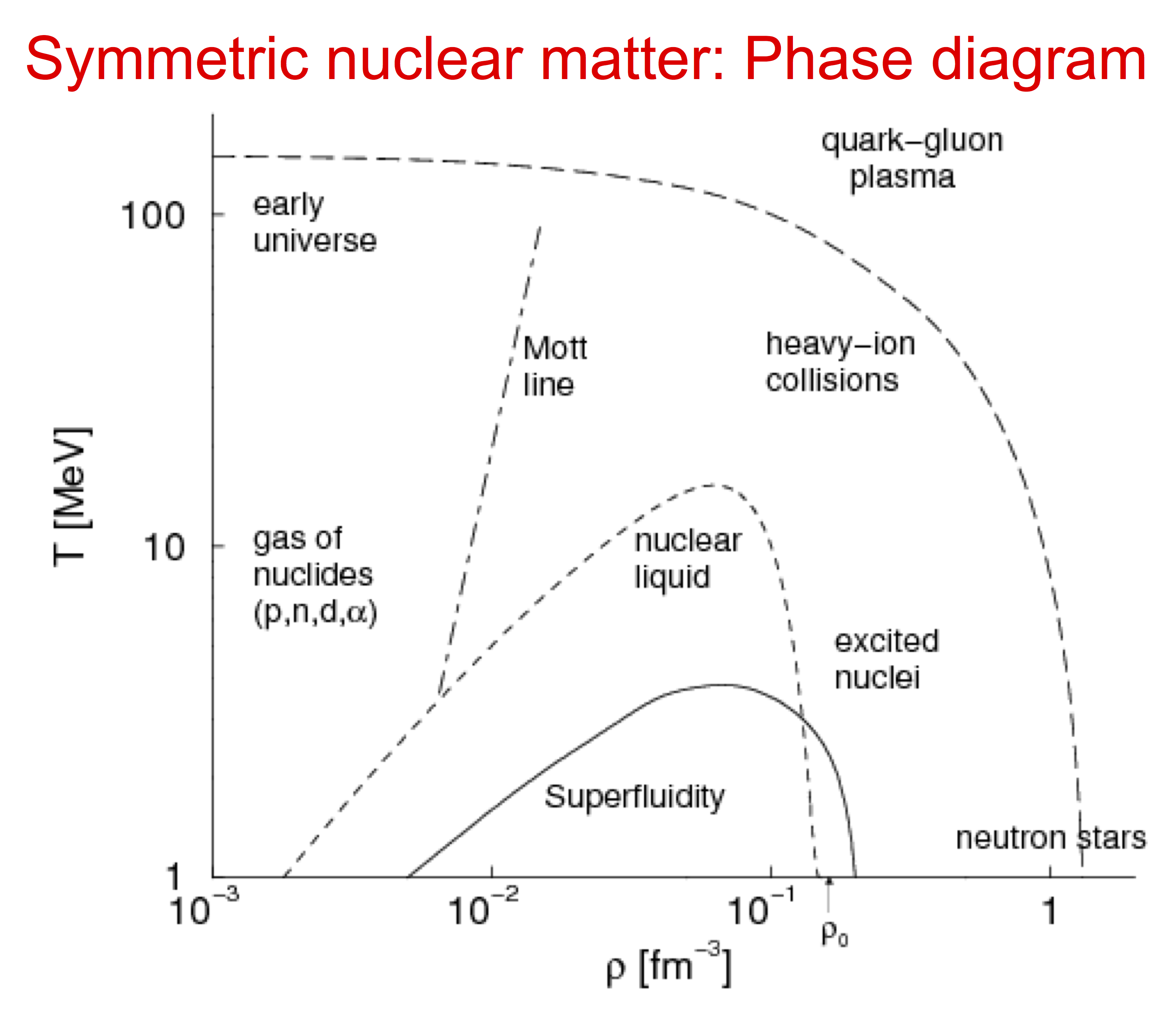}
   \end{subfigure}
   \quad
   \begin{subfigure}{0.4\textwidth}
     {Figure 1: Phase diagram of symmetric nuclear matter (schematic), showing the so-called Mott line as well as the liquid-gas like phase transition
     and the transition to the superfluid state. Presented at different talks since 1982, see also \cite{PD}.}
   \end{subfigure}
   \label{fig:PD}
 \end{figure}

It is a challenge to nuclear theory to describe the modification of the light-cluster properties caused by the surrounding
nuclear matter. The energies $E_i({\bf P})$ of the light clusters to be considered as quasiparticles with energies depending
not only on the c.o.m. momentum $\bf P$, but also on the densities of neutrons $n^{\rm tot}_n$, protons $n^{\rm tot}_p$ and on $T$.
The single nucleon states $n$, $p$ are treated along the same way.

In principle, one has to start from a many-body Hamiltonian. One has to calculate the spectral
function for the different channels, and sharp peaks can be interpreted as the corresponding quasi-particles.
Because we have no first-principle nucleon-nucleon forces such as the Coulomb force, the choice of the interaction potential
in the nuclear Hamiltonian remains a problem. Different parametrizations are obtained analyzing two-nucleon properties. We discuss 
some simple approximations in the following Section. However, to describe also three or four-nucleon systems, the pair potential
is no longer sufficient, and, for instance, density dependent forces are introduced.

Therefore it is advantageous to use directly the measured properties such as binding energies, rms radii, and scattering phase shifts,
avoiding the introduction of a potential. This way, the NSE can be formulated, but also the second virial coefficient can be expressed 
by measured scattering phase shifts. The properties of nuclear matter near saturation density are well reproduced by effective 
energy density functionals such as the Skyrme force or the relativistic mean-field (RMF) approach with empirical parameter values.
These single-nucleon quasiparticle energies contain already correlations beyond the Hartree-Fock (mean field) approximation.

To reproduce the EOS, quasiparticle energies for the light clusters can be introduced which depend on the thermodynamic 
variables $T, n^{\rm tot}_n, n^{\rm tot}_p$.
Such expressions have been introduced in Refs. \cite{Grigo1,SR} and are comprehensively discussed in \cite{Typel}.
They can be related to microscopic approaches for the single-nucleon self-energy and Pauli blocking discussed below.
As an alternative, the excluded volume concept is also used to mock the Pauli blocking term, see \cite{Hempel} and references given therein.
This empirical approach is not directly connected with the microscopic QS approach.


\section{\label{sec:shifts}Bound-state quasiparticle shifts}

Within a quantum statistical (QS) approach, the EOS $ n^{\rm tot}_\tau(T,\mu_n,\mu_p)$ is expressed in terms of the single-nucleon
spectral function which is related to the self-energy, see \cite{R,R2011}. Performing a cluster decomposition of the self-energy,
the relations
\begin{eqnarray}
\label{eos}
&&  n^{\rm tot}_n(T,\mu_n,\mu_p)= \frac{1 }{ \Omega} \sum_{A,\nu,P}N 
f_{A,Z}[E_{A,\nu}(P;T,\mu_n,\mu_p)] , \nonumber\\ 
&&  n^{\rm tot}_p(T,\mu_n,\mu_p)= \frac{1 }{ \Omega} \sum_{A,\nu,P}Z 
f_{A,Z}[E_{A,\nu}(P;T,\mu_n,\mu_p)] \, ,
\label{quasigas}
\end{eqnarray}
are obtained, where $\bf P$ denotes the center of mass (c.o.m.) momentum of the cluster (or, for $A=1$, the momentum of the nucleon). 
The internal quantum state $\nu$ contains the proton number $Z$ and neutron number $N=A-Z$ of the cluster, 
\begin{equation}
f_{A,Z}(\omega;T,\mu_n,\mu_p)=\frac{1}{ \exp [(\omega - N \mu_n - Z \mu_p)/T]- (-1)^A}
\label{vert}
\end{equation}
is the Bose or Fermi distribution function for even or odd $A$,
respectively, that is depending on $\{T,\mu_n,\mu_p\}$.

For the $A$-nucleon cluster, the  in-medium Schr\"odinger equation 
\begin{eqnarray}
&&[E_{\tau_1}(p_1)+\dots + E_{\tau_A}(p_A) - E_{A, \nu}(P)]\psi_{A \nu P}(1\dots A)
\\ &&
+\sum_{1'\dots A'}\sum_{i<j}[1-n(i)- n(j)]V(ij,i'j')\prod_{k \neq 
  i,j} \delta_{kk'}\psi_{A \nu P}(1'\dots A')=0\,\nonumber 
\label{waveA}
\end{eqnarray}
is derived from the Green function approach.
This equation contains the effects of the medium in the single-nucleon quasiparticle shift 
which is given, for instance, by the RMF expression \cite{Typel2005} ($n_B=n^{\rm tot}_n+n^{\rm tot}_p,\,\,Y_p=n^{\rm tot}_p/n_B$)
\begin{equation}
\label{DDRMF}
E_\tau(p;T,n_B,Y_p) = \sqrt{ \left[ m_\tau c^2-S(T,n_B,Y_p) \right]^2+\hbar^2 c^2 p^2} 
+ V_\tau(T,n_B,Y_p)- m_\tau c^2 .
\end{equation}
Approximation formulae for the RMF potentials $S, V$ are found, e.g., in \cite{Rarxiv}.

As well, the effects of the medium are obtained from the Pauli blocking terms given by the occupation numbers 
$n(1;T,\mu_n,\mu_p)$
in the phase space of single-nucleon states $|1 \rangle \equiv |{\bf p}_1,\sigma_1,\tau_1 \rangle$.  
The occupation numbers can be approximated by a Fermi distribution with effective parameter values for 
temperature and chemical potentials, see \cite{Rarxiv}.
Thus, two effects have to be considered, the quasiparticle
energy shift and the Pauli blocking. 

We obtain the cluster quasiparticle shifts 
\begin{equation}
\label{qushift}
E_{A,\nu}(P)-E^0_{A,\nu}(P)
=\Delta E_{A,\nu}^{\rm SE}(P)+\Delta E_{A,\nu}^{\rm Pauli}(P) 
+\Delta E_{A,\nu}^{\rm Coulomb}(P) 
\end{equation}
with the free contribution $ E^0_{A,\nu}(P)=E^0_{A,\nu}+\hbar^2 P^2/(2Am) $.
Expressions for the in-medium self-energy shift $\Delta E_{A,\nu}^{\rm SE}(P;T,n_B,Y_p) $ and Pauli blocking 
$ \Delta E_{A,\nu}^{\rm Pauli}(P;T_{\rm eff},n_B,Y_p) $ 
are given in \cite{Rarxiv}. 
The Coulomb shifts for the light elements with $Z \leq 2$ considered here are small compared with the 
other contributions and are omitted. 


The Pauli blocking contains the nucleon-nucleon interaction potential $V(ij,i'j')$ which has been taken in different approximations.
The main feature is the overlap of the cluster wave function in momentum space with the Fermi sphere so that an approximate form of 
the wave function can be used, reproducing characteristic parameters such as the rms radii. Simple expressions for the interaction potential
have been considered in \cite{R,R2011} and will be discussed in the following Section.

\begin{figure}[!hbt]
\begin{subfigure}{0.4\textwidth}
   \includegraphics[width=\textwidth]{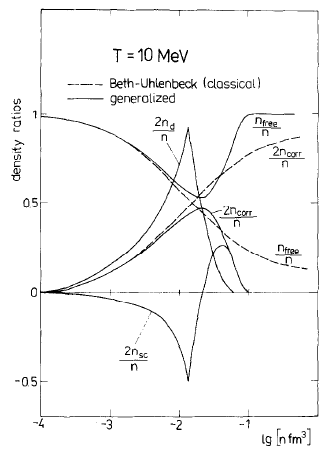}
   \end{subfigure}
   \quad
\begin{subfigure}{0.58\textwidth}
   {Figure 2: Two-particle correlations: Fraction of correlations as function of the baryon density $n$ for $T=10$.
     Results without in-medium corrections (dashed) are compared with results obtained from the generalized Beth-Uhlenbeck formula.
     Bound state contribution and the continuum contribution are also given. From \cite{SRS}.}
\end{subfigure}
   \label{fig:SRS}
 \end{figure}

 \section{\label{sec:Pauli}Interaction potentials and Mott line}
 
 As a first step, we consider only $A\leq 2$. The  generalized Beth-Uhlenbeck approach has been investigated by Schmidt {\it et al.}  \cite{SRS}.
 Results for the separable Paris (PEST) interaction potential, rank = 4, are given. 
 The result for the composition of symmetric matter at $T=10$ MeV is shown in Fig.~2.
 The sharp kink in the bound state contribution to the density at baryon density $n^{\rm Mott}_{d,{\rm Schmidt}}=0.013$ fm$^{-3}$ indicates the Mott line 
 where the bound states at ${\bf P}=0$ merge with the continuum. The total correlated density $n_{\rm corr}$ (containing also the contribution 
 of the continuum) is increasing up to densities of about 0.02 fm$^{-3}$ and then decreasing, vanishing at $n_B \approx 0.1$ fm$^{-3}$.
 
Calculations for a more extended parameter range in the $T-n_B$ plane have been performed by Stein {\it et al.} \cite{ZPhys}.
For simplicity a  rank = 1 Yamaguchi potential was used which is separable
and attractive only. It takes into account $S$-wave scattering
($c = ^1\!\!S_0, ^3\!\!S_1$) and depends on the relative momenta of
the incoming and outgoing two particles and the coupling
strength in the respective channel,
\begin{equation}
\label{Yama}
V^{\rm Y,L}_c(p, p') = -\lambda_c \frac{1}{p^2+\beta^2}  \frac{1}{{p'}^2+\beta^2}
\end{equation}
where $\beta = 1.4488$ fm$^{-1}$ is the inverse potential range,
$\lambda_{^1S_0}=2994$ MeV fm$^{-1}$ and $\lambda_{^3S_1}=4264$ MeV fm$^{-1}$ is the coupling
strength in the spin-singlet and in the triplet channel, respectively.
The parameters are fitted to the empirical nucleon-nucleon
scattering phase shifts and the vacuum bound state energy
of the deuteron ($E_d = -2.225$ MeV) which occurs in
the spin-triplet channel. The coupling of the $^3S_1$ to the $^3D_1$ channel
is neglected.

Results are shown in Fig.~3. For $T=10$ MeV, the Mott line has the value $n^{\rm Mott}_{d,{\rm Stein}}(T=10\, {\rm MeV}) = 0.02$ fm$^{-3}$
and the maximum contribution of correlated density occurs at 0.04 fm$^{-3}$, but a contribution of about 20 \% remains at saturation density.

We performed  calculations within the generalized Beth-Uhlenbeck approach \cite{SRS} for a simple separable Gaussian potential, 
\begin{equation}
\label{seppot}
 V^{\rm Y,G}_c(12,1'2')=-\lambda_c e^{-\frac{({\bf p}_1-{\bf p}_2)^2}{4\gamma^2}}e^{-\frac{({\bf p}'_1-{\bf p}'_2)^2}{4\gamma^2}}
 \delta_{\sigma,\sigma'}\delta_{\tau,\tau'}
\end{equation}
 with $\lambda_d= 1287.37$ MeV for the deuteron (isospin 0) channel, $\gamma = 1.474$ fm$^{-1}$, see \cite{R,R2011},
 adapted to binding energy and point rms radius of the deuteron. The Pauli blocking term for this potential has been evaluated 
 and parametrized in \cite{R2011}. For instance, for $T=10$ MeV the first order Pauli blocking shift is \cite{R} 352.5 MeV fm$^3 \times n_B$
 so that the Mott line $n^{\rm Mott}_{d,(1)}(T=10\, {\rm MeV}) = 0.0063$ fm$^{-3}$ follows. The exact solution gives $n^{\rm Mott}_{d,{\rm Y,G}}(T=10\, {\rm MeV}) = 0.009585$ fm$^{-3}$.
 
 \section{Different nucleon-nucleon potentials}
 
 To investigate the role of the hard-core repulsion in the nucleon-nucleon interaction, we considered a force of the  Mongan type \cite{Mongan}
\begin{equation}
\label{seppot}
 V^{\rm M}_c(12,1'2')=-\lambda_{c,a} e^{-\frac{({\bf p}_1-{\bf p}_2)^2}{4\gamma_a^2}}e^{-\frac{({\bf p}'_1-{\bf p}'_2)^2}{4\gamma_a^2}}+\lambda_{c,r} e^{-\frac{({\bf p}_1-{\bf p}_2)^2}{4\gamma_r^2}}e^{-\frac{({\bf p}'_1-{\bf p}'_2)^2}{4\gamma_r^2}}
 \delta_{\sigma,\sigma'}\delta_{\tau,\tau'}
\end{equation}
with $ \lambda_{d,a}=1645.89$ MeV and $\gamma_a=1.749$ fm$^{-1}$ for the attractive part and $ \lambda_{d,r}=445.843$ MeV and $\gamma_r=2.49$ fm$^{-1}$ for the repulsive part. Only a small decrease $n^{\rm Mott}_{d,{\rm M}}(T=10\, {\rm MeV}) = 0.0094967$ fm$^{-3}$ was obtained.

The use of a Yamaguchi nucleon-nucleon force with Lorentzian form factors (\ref{Yama}) gives nearly the same result for the Mott line,
 $n^{\rm Mott}_{d,{\rm Y,L}}(T=10\, {\rm MeV}) = 0.0095542$ fm$^{-3}$. 
 
 We conclude that Pauli blocking and the Mott line ($n^{\rm Mott}_{d,{\rm Y,L}},\,n^{\rm Mott}_{d,{\rm Y,G}},\,n^{\rm Mott}_{d,{\rm M}}$) are not very sensitive to the details of the nucleon-nucleon interaction potential. Global features such as the binding energy, the rms radius of the bound states as well as the scattering length and the low-$k$ expansion of the phase shifts are characteristics for the extension of the wave function in momentum space. The overlap with the single-particle distribution of the nuclear medium is relevant for the Pauli blocking effect.
 
 The main reason for the different results obtained above for the Mott line ($n^{\rm Mott}_{d,{\rm Schmidt}},\,n^{\rm Mott}_{d,{\rm Stein}}$) is due to the treatment of correlations evaluating the self-energy and Pauli blocking terms. If in Eq. (\ref{waveA}) the single-nucleon occupation $n(i)$ is replaced by the single-nucleon 
 distribution $f_{1,\tau}[E_\tau(p_i,T,\mu_n,\mu_p)]$ with the chemical potentials and temperature consistent with Eq. (\ref{quasigas}), only the free nucleons are taken into account for the phase occupation. The total baryon density $n_B$ is larger than the free nucleon density because of the contribution of the correlated density.
 
 It is necessary to take the contribution of correlations into account when calculating the mean-field and Pauli blocking shifts. The occupation number distribution $n(i)$ can be approximated by a Fermi distribution with parameter values $T^*,\mu_n^*,\mu_p^*$ describing the phase space occupation by free as well as correlated nucleons \cite{Rarxiv}. The Mott line obtained this way is $n^{\rm Mott}_d(T=10\, {\rm MeV}) = 0.0087987$ fm$^{-3}$.
 \hspace{-2cm}
\begin{figure}[!hbt]
   \begin{subfigure}{0.45\textwidth}
   \includegraphics[width=\textwidth]{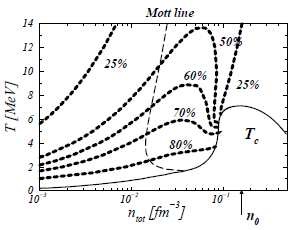}
   \end{subfigure}
   \quad
   \begin{subfigure}{0.54\textwidth}
     {Figure 3: Phase diagram of symmetric matter, with only two-particle correlations. Fraction of correlations 
     (bound state and continuum) to the density are given,
     as well as the Mott line and the transition temperatur $T_c$ for superfluidity. From \cite{ZPhys}.}
  \end{subfigure}
   \label{fig:Stein}
 \end{figure}
\section{\label{sec:Effect}Effective medium correction terms}

We give two examples for effective descriptions of the medium modification of the contributions of correlations.
We can introduce effective binding energies $B^{\rm eff}_i(T,n^{\rm tot}_n,n^{\rm tot}_p)$ depending only on the temperature $T$ 
and the total neutron/proton densities 
$n^{\rm tot}_n,n^{\rm tot}_p$, not on the c.o.m. momentum $\bf P$, and including excited states and the continuum of the channel $i$, 
using the definition
\begin{equation}
\label{effB}
 \sum_{\nu,P} 
f_{A,Z}[E_{c,\nu}(P;T,\mu_n,\mu_p)] =g_i  \sum_{P} f_{A,Z}[-B^{\rm eff}_i(T,n^{\rm tot}_n,n^{\rm tot}_p)+\frac{\hbar^2}{2 A m}P^2] .
\end{equation}
In the non-degenerate case, the sum over $P$ is easily performed, $g_i$ is the degeneration factor of the ground state.

Similar to the RMF expressions for the single-nucleon quasiparticle shifts, after an appropriate parametrization 
of  $B^{\rm eff}_i(T,n^{\rm tot}_n,n^{\rm tot}_p)$
a simple evaluation of the EOS (\ref{eos}) is possible. The Mott point $B^{\rm eff}_i(T,n^{\rm tot}_n,n^{\rm tot}_p)=0$ 
is a signature of medium effects and has been 
observed recently from measured yields in HIC \cite{Natowitz}. Theoretical approximations \cite{Grigo1,SR,Typel} 
$n_{B,d}^{\rm Mott \,point}(T=10\, {\rm MeV})=0.005225$ fm$^{-3}$, $n_{B,d}^{\rm Mott \,point}(T=5\, {\rm MeV})=0.00292$ fm$^{-3}$  are in 
reasonable agreement with the experimental data \cite{Natowitz} 
$n_{B,d}^{\rm Mott \,point}(T=4.5\, {\rm MeV})\approx 0.001$ fm$^{-3}$.

The other semi-empirical approach is the introduction of an excluded volume, see \cite{Hempel}. 
This parametrization of the medium effects is not very specific for the various constituents of the nuclear matter and demands some improvements
when it shell be related to the microscopic QS approach. However, it is rather efficient for exploratory calculations.

\section{\label{sec:Con}Conclusions}

The quasiparticle energies $E_i(P;T,n^{\rm tot}_n,n^{\rm tot}_p)$ of the element $i$ depends on the c.o.m. momentum $P$,
but also on the thermodynamic parameter values for $T,n^{\rm tot}_n,n^{\rm tot}_p$. Pauli blocking leads to the dissolution of bound states. 
In particular, $E_i(0;T,n^{\rm tot}_n,n^{\rm tot}_p)=0$ gives the so-called Mott line $T_{\rm Mott}(n_B)$ The correlated part
of the density, however, contains besides the contribution of bound also the contribution of scattering states.
Altogether, the correlated density $n_i(T,n^{\rm tot}_n,n^{\rm tot}_p)$ is a smooth function of  $T,n^{\rm tot}_n,n^{\rm tot}_p$. 

Calculations for the composition of nuclear matter have been performed for the contribution of two-nucleon correlations.
We found no strong dependence on the chosen nucleon-nucleon potential. Neglecting the details,
one can introduce an effective shift of the quasiparticle energies (effective binding energy $B^{\rm eff}_i$) which reproduces the contribution of two-nucleon correlations.
This is a simple concept to introduce the effect of cluster formation into the EOS.
For this effective shift, also a Mott point can be introduced where the effective shift compensates the binding energy of the cluster.

Experimental values have been obtained recently \cite{Natowitz} and have been compared with theories that use effective shifts \cite{Grigo1,SR,Typel}.
Good coincidence has been found. However, one should not identify the Mott point obtained there with the Mott line describing the 
dissolution of bound states at $\bf P$ = 0.

The dependence on temperature cannot be described by a simple excluded volume concept. According to \cite{R},
the deuteron Mott line $n^{\rm Mott}_d(T)$ for symmetric matter derived from the first order Pauli blocking shift has the following values: 
$n^{\rm Mott}_{d,(1)}(10\, {\rm MeV})=0.0063$ fm$^{-3}$,   $n^{\rm Mott}_{d,(1)}(20\, {\rm MeV})=0.0128$ fm$^{-3}$,  $n^{\rm Mott}_{d,(1)}(50\, {\rm MeV})=0.0385$ fm$^{-3}$, 
$n^{\rm Mott}_{d,(1)}(100\, {\rm MeV})=0.0968$ fm$^{-3}$,  $n^{\rm Mott}_{d,(1)}(140\, {\rm MeV})=0.154$ fm$^{-3}$. 
At very high temperatures deuterons can survive up to saturation densities,
as also observed recently at LHC experiments.

Similar approaches are possible also for other clusters with $A=3,4$ \cite{Rarxiv}. Because the clusters are more bound, 
the continuum contributions are less relevant. The use of the effective shift instead of the momentum-dependent quasiparticle shift (Mott line)
is better founded than in the deuteron case.

\vskip 10mm
\centerline{\bf Acknowledgment}
The author thanks David Blaschke for helpful discussions.
\vskip 10mm


\end{document}